**A simple multipurpose double-beam optical image analyzer**

A. Popowicz[1,a)] and T. Blachowicz[2]

[1]*Silesian University of Technology, Institute of Automatic Control, Akademicka str. 16, 44-100 Gliwice, Poland,*

[2]*Silesian University of Technology, Institute of Physics - Center for Science and Education, S. Konarskiego 22B str., 44-100 Gliwice, Poland*

In the paper we present a low cost optical device which splits the light in the focal plane into two separate optical paths and collimates it back into a single image plane, and where a selective information processing ca be carried out. The optical system is straightforward and easy implementable as it consists of only three lens and two mirrors. The system is dedicated for imaging in low-light-level conditions in which widely used optical devices, based on beam-splitters or dichroic mirrors, suffer from light loss. We expose examples of applications of our device, using a prototype model. The proposed optical system may be employed for: monitoring the objects located in different distances from observer (1), creating regions of different magnification within a single image plane (2), high dynamic range photometry (3), or imaging in two wavelength bands simultaneously (4).

[a)]Author to whom correspondence should be addressed. Electronic mail: adam.popowicz@polsl.pl.

# I. INTRODUCTION

Simultaneous observation of various optical properties of objects requires light splitting. It can be performed using either a beam splitter[1-4] or a dichroic mirror.[5-7] While the first one allows for spreading the light with a given proportion and in a wide wavelength range, the latter exposes wavelength dependent reflection characteristics. A triple-beam CCD camera, inveted by V.S. Dhillon et al.[8], can be a good example of such beam splitting device. This instrument incorporates dichroic mirrors which allow for observations in multiple bands for astrophysical purposes. As a matter of fact, the use of double beam astro-photometry has relatively long tradition.[9-11] A recent implementation of a beam splitting technique can be found in off-the-shelf product proposed by Hamamatsu, which allows for compact dual wavelength observations.[12] In fact, there are plenty of various multi-channel systems which are based either on beam splitters or on dichroic mirrors.[13-14] They can be also employed to observe scene with multiple focal points, thus obtaining sharp images of several separated objects.[15-16] The multiple focal plane microscopy is a prominent example of the imaging technique utilizing beam splitting.[16] A general idea of an optical system which divides the light, is presented in Fig. 1.

Since the mentioned image splitters create multiple copies of an image, the size of sensor has to be increased otherwise several cameras are required, which obviously increase the overall cost of such a system. Moreover, when one is interested in registering two (or more) objects which have to be observed through different filters (object A in F1 filter, object B in F2 filter, see Fig. 1.), the mentioned image splitters are not an optimal solution. It is because they actually divide a whole light beam, not the image itself, so that the light from all the objects is spread into multiple optical paths. Eventually, only a portion of light, emitted by a given object, is transferred through a dedicated filtering stage, while its remaining flux is virtually lost, hence, it is transferred unnecessarily to other optical paths.

In this paper we show a competitive concept of a simple optical device which splits the image plane into two separate optical paths and combines them back to a common focal plane. It is almost free of light loss, therefore it can be employed in low light level conditions, when several objects have to be observed using different optical treatment (filtering, polarizing, etc.). It also does not require additional imaging area since the size of field of view is preserved. Our constructed prototype allowed for verifying its capabilities and presenting some of its main applications. Several designing hints as well as possible extensions of the optical system are discussed below.

**II. DESIGN CONSIDERATIONS AND APPLICATIONS**

Our aim is to divide the image plane into two optical paths so that both image areas can be processed separately. In our device a certain part of the whole image is extracted for the secondary beam, while the remaining part stays in the primary beam. The scheme of the proposed system implementing this idea is presented in Fig. 2.

The device consists only of 5 elements (3 lenses and 2 mirrors) and employs spatial filtering of light within a focal plane of a telescope (or lens). There are two separate light paths, called direct (green) and indirect (red). Similarly, there are two focal planes: primary and secondary. While the former is created by a telescope, the image sensor is placed in the latter one. Since the two paths are well isolated, it is possible to apply arbitrary modifications to each of them, using, -e.g. filters, polarizers or changing optical magnification. The size of mirror M1 modifies the image area, which is transmitted through indirect path. The light from the remaining part of the primary focal plane is collimated by the lens L1 in direct path and creates a final image in secondary focal plane. The light from indirect path is initially collimated by L2, so that it is coherent, and then, it is reflected by the mirror M2 and collimated back on the image sensor by L3.

In our demonstrator, the mirror M1 (diameter equals 6 mm) is located centrally in the primary focal plane, employing a 1.5 mm stick, which can slightly hide some details of final image. Fine positioning of all the elements is enabled by using adjustable holders and trails. We employed basic single-element, double-convex lenses, (50 mm focal length, f/1). A custom housing allowed for attaching full frame ATIK 11000 astronomical camera with Kodak KAI11000 interline CCD sensor installed. The whole device as well as the telescope (refractor, 80 mm aperture, 600 mm focal length) are mounted on a common optical rail. The relative position of telescope and optical board can be adjusted by regulated stands. The detailed views on the prototype is given in Fig. 3, while an exemplary image registered by the camera is shown in Fig. 4.

The presented prototype is definitely not free from either optical aberrations (especially coma) or vignetting, however, it was developed only as a concept demonstrator using the simplest optical elements. Importantly, the device was constructed as a tradeoff between the needs, requirements and currently owned elements. With a more sophisticated workshop, the optical system can be constructed with virtually no optical distortions and in a miniaturized version, as off-the-shelf device.

Proper design of the proposed optical path may be mechanically and optically challenging. During creation of our prototype we encountered numerous problems with positioning of the elements. Not only the light paths may overlap, but also the sizes of optical elements and holders introduce some difficulties.

Therefore, based on our experience with the prototype, we provide several constraints important for choosing right lenses and for keeping the required distances. As all necessary information can be easily derived from simple trigonometric calculations, only the most important tips and indications are listed below.

1. The diameter of L1 is critical for avoiding unwanted vignetting in the direct path. The elements with lower focal ratios will generally provide better performance. Unfortunately they reduce the length of the path making the indirect path harder to design.
2. ~~2.~~ To preserve the aspect ratio of the primary focal plane, the lens L1 should be placed exactly between both primary and secondary planes, in a distance $2 \cdot f_{L1}$ from each one. This is also the shortest path between the planes which minimizes the width of a whole instrument.
3. The size of L2 should be adjusted according to the size of M1 mirror and it will be always much smaller than L1.
4. Lens L3 should be at least the same size as L2. However, to avoid vignetting in the indirect path, we suggest using slightly larger L3.
5. The both paths should be kept well apart so that the light from any path is not transferred by the lenses installed in the other path.
6. The inclination angle of the light directed from the indirect path to the sensor has to be the smallest possible to avoid image deformations. Additionally, it should be minimized since the sensitivity of sensors decreases as the light comes from not perpendicular direction.

Finally, when the device is ready, it has to be properly calibrated. We found the following useful steps:

1. Direct the telescope toward a distant object.
2. Obtain a sharp image of an object within a plane of mirror M1 (primary focal plane) by adjusting telescope position.

3. Block the light in indirect path and adjust the position of L1 so that the sensor produces sharp images.
4. Block the light in the direct path, then remove the lens L3, and place a bright spot of light coming from L2 in the center of sensor's plane.
5. Mount lens L3 and adjust both lenses (L3 and L2) in the indirect path to achieve a desired magnification/sharpness level.

The presented device has several advantages over widely-used optical devices based on semi-transparent beam splitters or dichroic mirrors. These solutions allow for observing multiple copies of an object in different wavelength bands or with various magnifications. Unfortunately, such splitting has several drawbacks, listed below, which are not present or can be mitigated in our concept.

1. The creation of multiple copies of observed objects puts much higher demands on the size of image sensor, requiring sometimes several, well-synchronized cameras, for its proper operation. In our solution, the original (i.e. primary) field of view remains unchanged.
2. Each copy of an image has reduced intensity. In proposed device, the original intensity of each image part is very well preserved.
3. The light dividing using the beam splitters is strongly dependent on the angle of incidence, producing flat-fields distortions. This effect is also present in our concept, however only at the edges of image region transferred by the indirect path. Moreover, this problem can be reduced by changing orientation of M1 mirror, so that the light in the indirect path is reflected slightly back, toward the telescope.

The proposed device has, however, some disadvantages. There will be always some diffraction-related effects produced by the edges of M1 mirror. Additionally, the mirror, in our case, was installed on a stick, which hides some details of background image. Another noticeable problem appears due to the non-perpendicular orientation of indirect light beam, relatively to the sensor's plane. Although we did not noticed significant image distortions when the magnification of both paths was the same, it may be an issue when either the mirror M1 is larger or a greater magnification in indirect path is applied. Also, the sensitivity of image sensor decreases for light coming from not perpendicular directions. Therefore, its characteristics have to be carefully checked prior to the installation. In our case, the CCD had to be rotated by 90 degree, since the pixels exposed different sensitivity characteristics for vertical and horizontal directions, respectively.[17]

The proposed image splitting device can be easily extended by creating more than two optical paths. This can be accomplished by constricting a custom mirror M1 which directs the light into separate tracks or by inputting multiple M1-like mirrors, one next to another. One can imagine even a more sophisticated device which is capable of steering the positions of M1 mirrors, so that it is possible to precisely select each region of interest. The implementation of such multi-paths system using beam splitters would suffer from significant light loss, since the beam would have to be divided many times before it reaches a desired optical stage (i.e. filters, polarizers or focusing optics).

The splitting of paths allows for adjusting the optical magnification for both regions individually, which can create a zoomed region, as presented in Fig. 5. For such a purpose only a part of M1 area should reflect the light, so that the magnified region does not expand too much, overlapping with the remaining, background image. It can be obtained by creating custom mirror or by placing a diaphragm behind M1. The latter solution, employing rectangular blocking diaphragm, was utilized by us for obtaining the images from Fig. 5.

Obviously, the magnified region will be dimmer, since the light is spread over larger area. This effect may be compensated by placing natural density filter in direct optical path. This application of our device can be utilized when both smallish and much larger objects are observed in one field.

Another application of presented image splitter involves observations of objects, which require multiple focal points. For such a purpose, the lens L2 and L3 should be co-adjusted to create a sharp image of close objects, while the distance background is focused by the direct path. In an exemplary registration, presented in Fig. 6, we were able to observe details of brick chimney, located about 10 m from the telescope, while the focal point of direct path was set to the distant trees (approx. 1000 m away). Such configuration of proposed device can be employed to increase the resolving power for nearby objects.

When the photometry of separable objects has to be performed, the presented device can be utilized to enhance the dynamic range of measurements. This problem appears in the astronomical observations from the ground, wherein the reference (constant) star should be located very close, within so-called isoplanatic angle of the order of several arc seconds, to the measured star.[18,19] This requirement is dictated by the high correlation of scintillation effects for angularly close objects. Unfortunately the nearby stars are usually of much different magnitude, resulting in either saturating the image sensor (brighter reference) or creating only very little reference signal (dimmer reference). Since the stars are point sources, it is possible to apply the presented splitting device to blur the brighter star (in indirect path), while keeping the dimmer star in focus (in direct path). Thanks to such a position dependent blurring, all the photon counts are registered from both objects, making the photometry optimal. A somewhat similar approach was used in CoRoT satellite[20] to overcome the problem of sensor saturation. Herein, two optical channels were employed with two separate CCDs located at different distances from the focal point, to obtain two permanent degrees of star defocussing. In our solution, such selective blurring can be easily tuned. We are currently working on a high-quality and compact prototype of image splitting device for such an astronomical purpose.

Many other modifications of the optical paths in the proposed system can be applied. They may include inputting photometric filters or polarizes individually in each path, so that various optical properties of objects are observed. The presented concept can be also employed in numerous applications, for which the current beam splitting devices are used.

## III. CONCLUSIONS

In the paper we presented a simple optical device, which divides the light beam from the primary focal plane and collimates it back in the secondary focal plane. The proposed device can be employed for observations of various optical properties of multiple objects independently. In contrast to widely used systems based on beam splitters and dichroic mirrors, in our proposition the light from a given object is processed only by a required optical stage, thus no light is lost. It predestinates the device for a wide range of low-light-level applications.

The constructed prototype allowed for gaining the knowledge about the main advantages, drawbacks and construction challenges of the device. It also enabled showing some of its applications. With a more sophisticated mechano-optical workshop and with a use of high-quality elements, the system may be virtually free from vignetting and aberration problems which appeared in our demonstrator. It may be miniaturized to work as a compact, multi-purpose optical device.

**ACKNOWLEDGEMENTS**

Adam Popowicz was supported by Polish National Science Center, grant no. 2013/11/N/ST6/03051: Novel Methods of Impulsive Noise Reduction in Astronomical Images.

The work of the second author was partially supported by the Institute of Physics – CSE local project BK-243/2016.

The research was performed using the infrastructure supported by POIG.02.03.01-24- 099/13 grant: GeCONiI – Upper Silesian Center for Computational Science and Engineering.

Figure captions

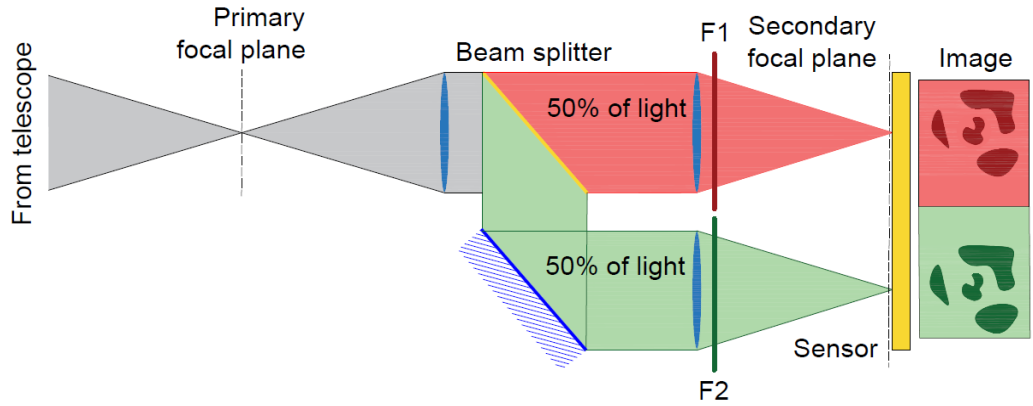

FIG. 1. A general idea of standard optical system with beam splitting. The light after collimation is subdivided by a 50/50 beam splitter. The final image is composed of two copies of an image from the primary focal plane. The splitted beam can be additionally passed through F1 and F2 spectral filters.

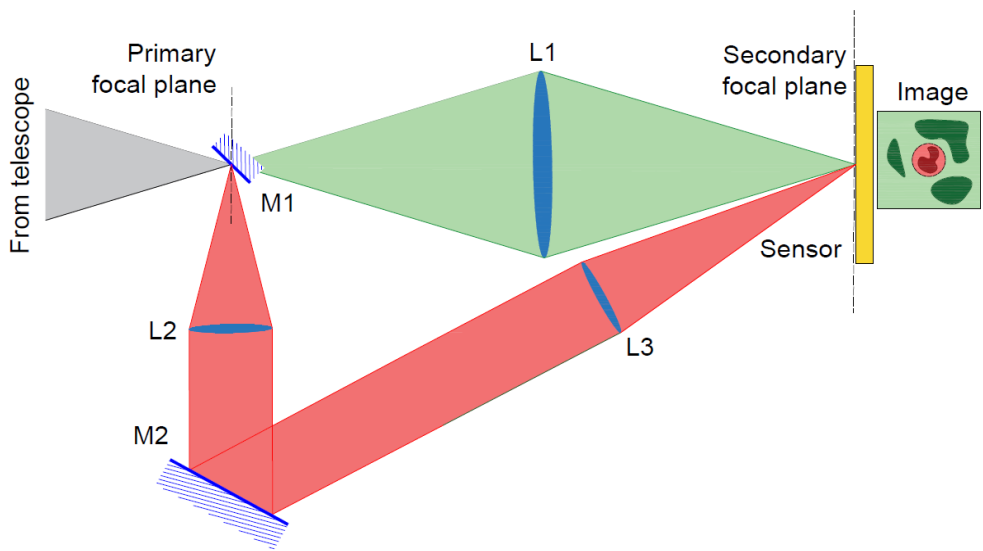

FIG. 2. Setup of the proposed double-beam optical image analyzer. The central part of primary focal plane is redirected by a small, circular mirror M1 into the indirect optical path

(red) and collimated with a use of L2 and L3 lenses. The remaining part is collimated in direct path (green) using L1 lens. The symbolic green and red annotated regions in a final image correspond the light transferred by direct and indirect path, respectively.

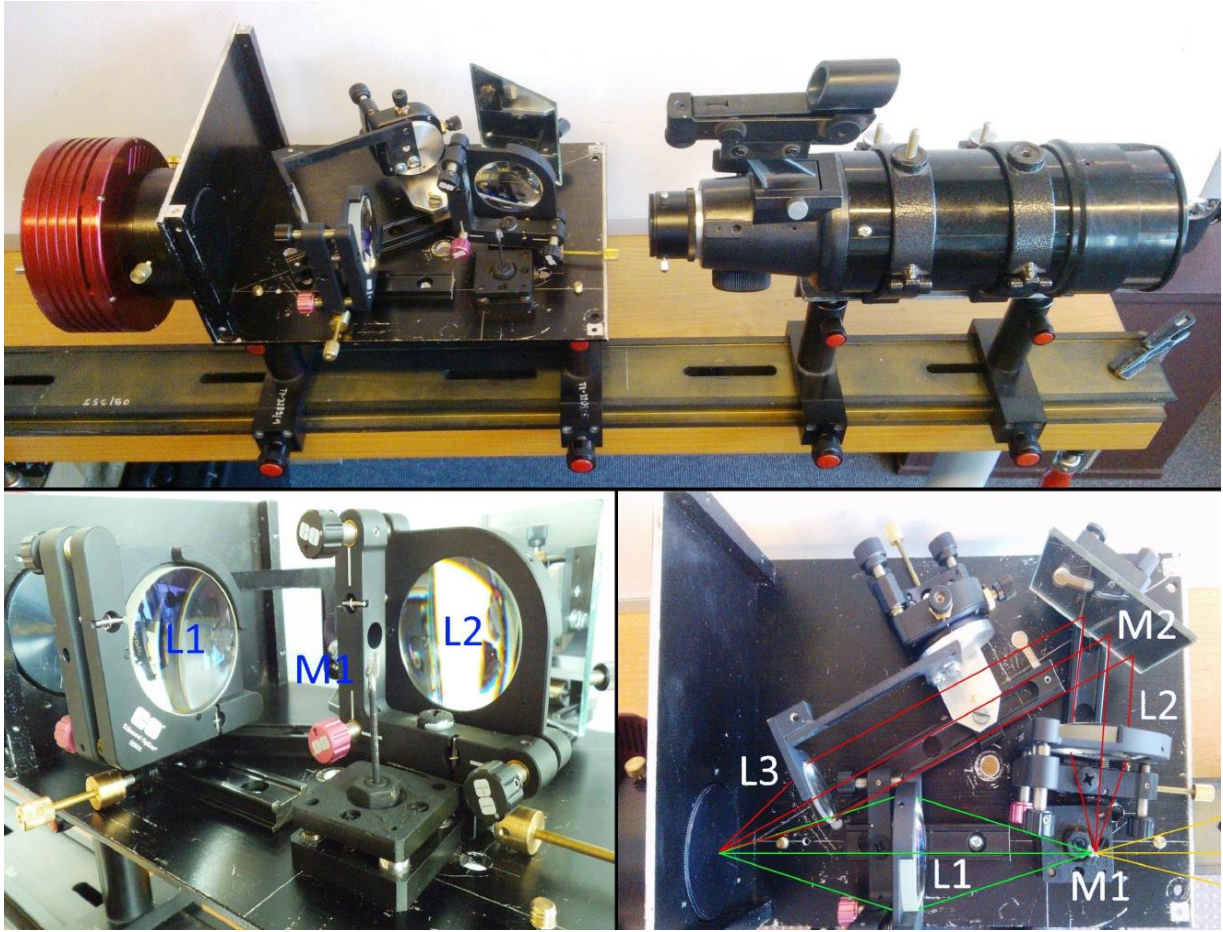

FIG. 3. Constructed optical path with camera and telescope attached. The presented images expose the device uncovered. Lower right image shows the utilized 6 mm M1 mirror with neighboring lens L1 (on left) and L2 (on right), while the lower left image presents an upper view with the light traces indicated (green - direct path, red - indirect path, yellow - light from telescope). In all images, the camera has the CCD sensor protected by a cover.

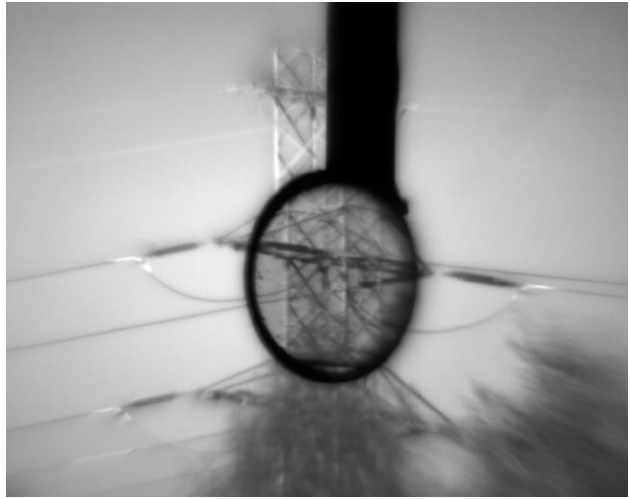

FIG. 4. An image obtained with the ~~a~~ proposed optical system showing a distant transmission tower. Slight dimming of the central region appears due to the decrease of pixel sensitivity for non perpendicular incident light angle. Vignetting and comma-related distortions emerge~~s~~ toward the image edges

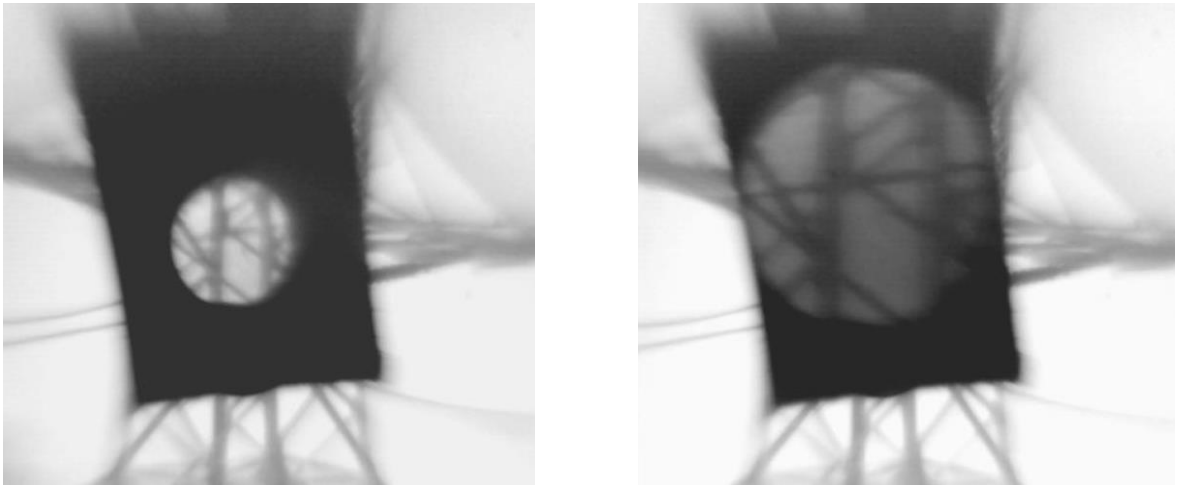

FIG. 5. Application of the optical path for tunable magnification of central image region: (a) no magnification, (b) 2x magnification applied. Black rectangle corresponds to a diaphragm placed behind the small mirror M1.

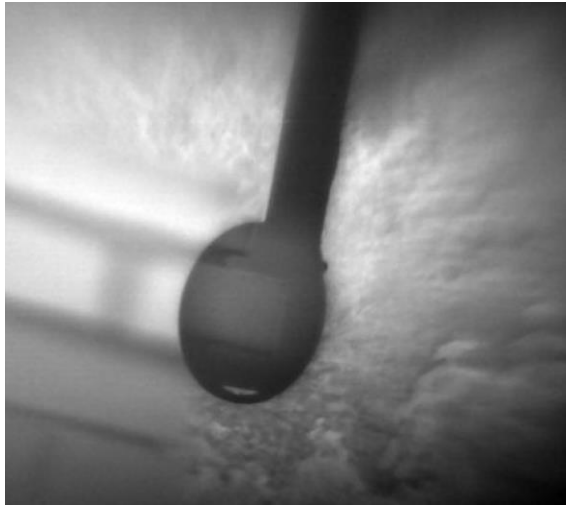

FIG. 6. Application of the optical path for observations in double-focus mode. The focus of direct optical path was adjusted for distant trees (~1000 m from an observer), while the indirect path was modified to register nearby brick chimney (10 m from the telescope).